\documentclass[a4paper,dvips,12pt]{article}
\usepackage{amsmath,latexsym,amssymb,epsfig,psfrag}
\usepackage{comment}
\usepackage{cite}
\usepackage{wasysym}

\newcommand{\be}{\begin{equation}}
\newcommand{\ee}{\end{equation}}

\newcommand{\de}{\partial}

\newcommand{\zero}{{{}_{(0)}}}
\newcommand{\ppi}{{{}_{(\pi)}}}

\makeatletter
\@addtoreset{equation}{section}
\makeatother

\begin{document}

\baselineskip=16pt
\begin{titlepage}
\begin{center}
\hfill{{\bf UAB-FT-575}}

\hfill{{\bf IPPP/04/16}}

\hfill{{\bf DCTP/04/32}}

\hfill{{\bf IFIC-04-60}}

\vspace{0.5cm}

\large {\sc \Large Supersymmetry from Boundary Conditions}

\vspace*{5mm}
\normalsize

{\bf G.~v.~Gersdorff~\footnote{gero@pha.jhu.edu},
L.~Pilo~\footnote{luigi.pilo@pd.infn.it},
M.~Quir\'os~\footnote{quiros@ifae.es},
A.~Riotto~\footnote{antonio.riotto@pd.infn.it},
V.~Sanz~\footnote{veronica.sanz-gonzalez@durham.ac.uk}}

\smallskip
\it{~$^{1}$~Department of Physics and Astronomy, Johns Hopkins University}\\
\it{Baltimore, MD 21218}

\smallskip
\medskip
\it{~$^{2,\,4}$~Department of Physics and INFN, Sezione di Padova \\
via Marzolo 8, I-35131 (Padova) Italy}

\smallskip
\medskip
\it{~$^{3}$~Instituci\'o Catalana de Recerca i Estudis Avan\c{c}ats (ICREA)}\\
\it{and}\\
\it{Theoretical Physics Group, IFAE/UAB}\\
\it{E-08193 Bellaterra (Barcelona) Spain}

\smallskip
\medskip
\it{~$^{5}$~Departament de F\'\i sica Te\` orica, IFIC,
Universitat de Val\` encia-CSIC} \\ \it{E-46071 Val\` encia, Spain}

\vskip0.3in
\end{center}

\centerline{\large\bf Abstract}

\noindent
We study breaking and restoration of supersymmetry in five-dimensional
theories by determining the mass spectrum of fermions from their
equations of motion. Boundary conditions can be obtained from either
the action principle by extremizing an appropriate boundary action
(interval approach) or by assigning parities to the fields (orbifold
approach). In the former, fields extend continuously from the bulk to
the boundaries, while in the latter the presence of brane mass-terms
cause fields to jump when one moves across the branes. We compare the
two approaches and in particular we carefully compute the non-trivial
jump profiles of the wavefunctions in the orbifold picture for very
general brane mass terms.  We also include the effect of the
Scherk-Schwarz mechanism in either approach and point out that for a
suitable tuning of the boundary actions supersymmetry is present for
arbitrary values of the Scherk-Schwarz parameter.  As an application
of the interval formalism we construct bulk and boundary actions for
super Yang-Mills theory. Finally we extend our results to the warped
Randall-Sundrum background.

\vspace*{2mm}

\end{titlepage}

\section{\sc Introduction}

A common feature of five-dimensional (5D) supersymmetric models are
fermions propagating in the bulk of the extra dimension. In order to
extract physical predictions at low energies, the four dimensional
mass spectrum of those fermions has to be known.  For instance
supersymmetry breaking is determined by the mass spectrum of the
gravitino and the existence of a zero mode signals an unbroken
supersymmetry. Similarly, when gauge multiplets propagate in the bulk
supersymmetry breaking is intimately linked to the existence of
gaugino zero modes. In particular if supersymmetry breaking is
implemented by non-trivial twist conditions, or Scherk-Schwarz (SS)
mechanism~\cite{ss}, it acts in the same way both in the gravitino and
the gaugino sectors.

The aim of this paper is to study fermions propagating in a
five-dimensional space-time, with coordinates $(x^\mu,y)$, where the
compact fifth dimension (with radius $R$) has two distinguished
four-dimensional hypersurfaces located at $y=0$ and $y=\pi R$.  Often
this space is constructed as the orbifold $S^1/\mathbb Z_2$,
identifying points on the circle related by the reflection of the
fifth coordinate $y\to -y$; in this approach $y=0$ and $y=\pi R$ are
fixed points and the resulting spacetime is a singular space without
boundaries.  Fields with odd parity with respect to the $\mathbb Z_2$
reflections are zero at the fixed points, while the normal derivative
of even fields is forced to vanish there.  The treatment of fermions
is complicated in the presence of brane actions localized at the
boundaries.  In the orbifold approach these brane actions are
introduced with delta-function distributions peaked at the location of
the orbifold fixed points.  The latter induce discontinuities in the
wave functions of the fermions, which take different values at the
fixed points and infinitesimally close to
them~\cite{Bagger:2001qi,Delgado:2002xf}. A possible way to avoid
these problems is working from the very beginning in the fundamental
region of the orbifold $[0, \pi R]$ and giving up the rigid orbifold
boundary conditions (BC's)~\footnote{With some abuse of language we
call boundary conditions the parity assignment for fields in the
orbifold in order to make contact to the interval approach.}: the
fields are then continuous and the BC's are determined by the action
principle applied to the bulk and boundary conditions. This is called
the {\em interval} approach: contrary to the orbifold approach the
spacetime is not singular but it has two boundaries at $y=0$ and
$y=\pi R$.  The interval approach leads to physically equivalent
results to those of the orbifold approach without any need of using,
as the latter, singular functions~\footnote{The interval approach is
sometimes called ``downstairs'' approach while the orbifold approach
is called ``upstairs'' approach.}.  To summarize in the orbifold
approach one imposes fixed (orbifold) BC's while the brane action
induces jumps, whereas in the interval approach one imposes continuity
and the brane action induces the BC's.

In a previous letter~\cite{us} we followed the interval approach and
showed how one can obtain consistent BC's by the use of the action
principle~\footnote{For an alternative approach
see~\cite{csfermion}.}: variation of the bulk action gives rise both
to a bulk term and a boundary term, while the variation of the
boundary action is always localized at the branes. Imposing the whole
variation to vanish the bulk terms give the usual bulk equations of
motion (EOM) while the boundary terms give rise to the corresponding
BC's (see~\cite{Csaki:2003dt} for a recent application to symmetry
breaking). In~\cite{us} it was also shown that for consistent
(nontrivial) BC's one needs to appropriately constrain the boundary
action; in particular, a vanishing brane action leads to inconsistent
BC's. The BC's can be seen to represent a point in the Riemann sphere
and hence are given by complex numbers $z_f$, $f=0,\pi$.  Explicit
formulae for mass eigenvalues and eigenfunctions as functions of $z_f$
can then be obtained. Whenever the BC's at the two branes are the
same, $z_0=z_\pi$, a zero mode exists and $N=1$ supersymmetry remains
unbroken.  An interesting phenomenon occurs when in addition a
Scherk-Schwarz~\cite{ss} breaking is turned on. The latter can be
implemented as a vacuum expectation value (VEV) for the auxiliary
vector field $\vec V_M$ which gauges the $SU(2)_R$ automorphism
symmetry~\cite{sshos} as $\langle \vec V_M\rangle=\delta_{M5}\,
\omega\, \vec q$, characterized by the SS parameter $\omega$ and the
SS direction (unit vector) $\vec q$. Whenever $\vec q$ is
aligned~\footnote{See below for a precise definition of this term.}
with either $z_0$ or $z_\pi$ , the mass spectrum becomes independent
on the SS parameter $\omega$. In particular this means that if in
addition $z_0=z_\pi$, supersymmetry remains unbroken whatever the
value of the SS parameter $\omega$ is. This was dubbed {\em persistent
supersymmetry} in Ref.~\cite{us}.  As an application of the interval
formalism, we write down the action of super Yang-Mills (SYM)
theory. A boundary action is required both for consistent BC's as well
as (global) supersymmetry.

In the present work we also shed some light on the relation of the
interval approach and the more common orbifold picture in which BC's
are fixed by the parity assignments of the fields. We show how in the
latter the mass spectrum can be computed by calculating the highly
generally nontrivial jump profiles across the branes. In order to
consistently treat the singularities arising from the presence of
delta-functions we formally use regularized delta-functions.  The
jumps determine the values of the wave functions an infinitesimal
distance away from the branes and thus can be used as new BC's in the
general formulae for the solution to the bulk EOM.  The impact of
fermionic brane mass terms on the spectrum as well as its relation to
the SS mechanism have been discussed before in a number of
papers~\cite{Bagger:2001qi,various,SSBC,Delgado:2002xf}, with
sometimes contradictory results due to incorrect treatment of the
discontinuities. We believe that our present paper gives a clear and
consistent treatment of this slightly involved issue by employing the
most general brane Lagrangian considered so far.

The structure of the paper is as follows: section~\ref{interval}
contains the results derived in~\cite{us} on the interval approach. In
section~\ref{SYM} the action of the Super Yang-Mills (SYM) theory is
derived. In section~\ref{orbifold} we give the treatment of the most
general boundary mass terms in the orbifold picture, as outlined
above.  Emphasis is put on the careful (regularized) treatment of the
singular profiles of the wave functions.  In section~\ref{warped} we
extend our results to the case of warped (RS) geometry. Finally in
section~\ref{conclusions} we draw our conclusions.

\section{\sc Fermions in the interval}
\label{interval}
In this section we recall some of the results of~\cite{us}.  We will
take the fermions to be symplectic-Majorana spinors, although a very
similar treatment holds for the case of fermionic matter fields
associated to Dirac fermions.  In particular we will consider the
gaugino case, the treatment of gravitinos being completely analogous.
The 5D spinors $\Psi^i$ satisfy the symplectic-Majorana reality
condition and we can represent them in terms of two chiral 4D spinors
according to~\footnote{We use the Wess-Bagger convention~\cite{wb} for
the contraction of spinor indices.}
\begin{equation}\label{uno}
\Psi^i = \begin{pmatrix} \eta^i_\alpha \\
\bar{\chi}^{i\,\dot{\alpha}} \end{pmatrix} \, , \qquad
\bar{\chi}^{i \,\dot{\alpha}} \equiv \epsilon^{ij} \,
\left(\eta^j_\beta \right)^\ast \,  \epsilon^{\dot{\alpha}
\dot{\beta}}
\; ,
\end{equation}
where $\epsilon_{ij}= i \,(\sigma_2)_{ij}$ and $\epsilon^{im}
\epsilon_{jm} = \delta^i_j$.  Consider thus the bulk Lagrangian
\begin{equation}
{\cal L}_{\text{bulk}} =i\, \bar{\Psi} \gamma^{{}_M}
D_{{}_M} \Psi =\frac{i}{2} \bar{\Psi} \gamma^{{}_M}
D_{{}_M} \Psi - \frac{i}{2} D_{{}_M} \bar{\Psi} \gamma^{{}_M} \Psi \; ,
\end{equation}
where the last equation is not due to partial integration but holds
because of the symplectic-Majorana property, Eq.~(\ref{uno}). The
derivative is covariant with respect to the $SU(2)_R$ automorphism
symmetry and thus contains the auxiliary gauge connection $V_M$.  The
field $V_M$ is non propagating and appears in the off-shell
formulation of 5D supergravity~\cite{zucker}.  A vacuum expectation
value (VEV)~\footnote{Consistent with the bulk equation of motion
$d\,(\vec q\cdot \vec V)=0$~\cite{zucker}.}
\be V_{_M} = \delta^5_{_M} \, \frac{\omega}{R} \,
\vec{q}\cdot\vec\sigma \, , \qquad {\vec{q}}^{\, 2}=1
\label{SSVEV}
\ee
implements a Scherk-Schwarz supersymmetry breaking mechanism~\cite{ss}
in the Hosotani basis~\cite{hos,sshos}.  The standard form of the SS
mechanism, originally introduced for circle compactification, can be
recovered by a gauge transformation $U$ that transforms away $V_{_M}$
but twists the periodicity condition for fields charged under
$SU(2)_R$ on the circle.  As a matter of fact, in the interval a SS
breaking term is equivalent to a suitable modification of the BC's at
one of the endpoints.  The unitary vector $\vec{q}$ points toward the
direction of SS breaking.  We supplement the bulk action by the
following boundary terms at $y=y_f$ ($f=0,\pi$) with $y_0=0$ and
$y_\pi=\pi R$
\begin{equation}
{\cal L}_{f} = \frac{1}{2} \bar{\Psi} \left(T^{{(f)}} + \gamma^5 \,
V^{{(f)}} \right) \Psi =\frac{1}{2} \,{\eta^i}M^{_{(f)}}_{ij} \eta^j
+{\rm h.c.}\; ,
\label{bmasses}
\end{equation}
where $T^{{(f)}}$ and $V^{{(f)}}$ are matrices acting on $SU(2)$
indices,
\begin{equation}
M^{{(f)}}=i\sigma_2 \, (T^{{(f)}}-i V^{{(f)}})
\label{TVm}
\end{equation}
and we have made use of the decomposition (\ref{uno}).  Notice that
the mass matrix is allowed to have complex entries.  Without loss of
generality we take it to be symmetric, which enforces $T^f$ and $V^f$
to be spanned by Pauli matrices.

The total Bulk + Boundary action is then given by
\begin{equation}
S= S_{\text{bulk}}+S_{\text{boundary}}=\int d^5x \, {\cal
L}_{\text{bulk}} + \int_{y =0} d^4x \,{\cal L}_{0} - \int_{y =\pi R}
d^4x \,{\cal L}_{\pi} \quad .
\label{total}
\end{equation}
The variation of the total action gives the standard Dirac-like bulk
equation of motion provided that all the boundary pieces vanish. The
latter are given by
\begin{equation}
\left.\left[\delta \eta^i \left( \epsilon_{ij} +
M_{ij}^{{}_{(f)}}\right) \eta^j + \, \text{h.c.}  \right]\right|_{y
= y_f} = 0 \; . \label{bf}
\end{equation}
Since we are considering unconstrained variations of the fields, the
BC's we obtain from Eqs.~(\ref{bf}) are given by
\begin{equation}
\left.\left( \epsilon_{ij} + M_{ij}^{{}_{(f)}}\right) \eta^j
\right|_{y = y_f} = 0 \; . \label{bf2}
\end{equation}
These equations only have trivial solutions (are over-constrained)
unless
\be
\det\left(\epsilon_{ij} + M_{ij}^{{}_{(0)}}
 \right) =\det\left( \epsilon_{ij} + M_{ij}^{{}_{(\pi)}}\right)=0\;.
\label{det}
\ee
Imposing these conditions we get the two complex BC's which are needed
for a system of two first order equations.  Note that this means that
an arbitrary brane mass matrix does not yield viable BC's; in
particular a vanishing brane action is inconsistent~\footnote{In the
sense that the action principle does not provide a consistent set of
BC's as boundary equations of motion.} since $\det(\epsilon_{ij})\neq
0$~\footnote{Notice that this agrees with the methods recently used in
Ref.~\cite{moss}.}.  However this does not imply that the familiar
orbifold BC's $\eta_1=0$ ($\eta_2=0$) can not be achieved; in the
interval approach they correspond to $M=\sigma^1$ ($M=-\sigma^1$).

The BC's resulting from Eqs.~(\ref{bf2}) are of the form
\begin{equation}
\left.\left(c^1_f \, \eta^1 + c^2_f \, \eta^2 \right) \right|_{y =y_f}
= 0 \quad ,
\label{cbc}
\end{equation}
where $c^{1,2}_f$ are complex parameters or, setting $z_f =-
(c^1_f/c^2_f)$
\begin{equation}
\left.\left(\eta^2 -z_f \, \eta^1 \right) \right|_{y =y_f} = 0, \quad z_f\in
\mathbb C\quad .
\label{bcf}
\end{equation}
Physically inequivalent BC's span a complex projective space
$\mathbb{C}P^1$ homeomorphic to the Riemann sphere.  In particular
$z_f =0$ leads to a Dirichlet BC for $\eta_2$, and the point at
infinity $ z_f =\infty$ leads to a Dirichlet BC for $\eta_1$.  Notice
that these BC's come from $SU(2)_R$ breaking mass terms. Special
values of $z_f$ correspond to cases when these terms preserve part of
the symmetry of the original bulk Lagrangian. In particular when both
the SS and the preserved symmetry are aligned those cases can lead to
a persistent supersymmetry as we will see.  Once (\ref{det}) is
satisfied the values of $z_f$ in terms of the brane mass terms are
given by
\be
z_f=-\frac{M_{11}^{_{(f)}}}{1+M_{12}^{_{(f)}}}
 =\frac{1-M_{12}^{_{(f)}}}{M_{22}^{_{(f)}}}
\label{zBC}
\ee
where the second equality holds due to condition (\ref{det}).

The mass spectrum is found by solving the EOM with the BC's
(\ref{bcf}).  To simplify the bulk equations of motion it is
convenient to go from the Hosotani basis $\Psi^i$ to the SS one
$\Phi^i$, related by the transformation
\begin{equation}
\Psi = U \, \Phi ,\quad U=
\exp{\left(-i\, \vec{q} \cdot \vec{\sigma}\, \omega\, \frac{y}{R}\right)}  \;.
\label{cb}
\end{equation}
In the SS gauge the bulk equations read
\begin{equation}
i \, \gamma^{{}_M} \de_{{}_M} \Phi = 0 \quad .
\end{equation}
We now decompose the chiral spinor $\eta^i(x,y)$ in the Hosotani basis
as $\eta^i(x,y) = \varphi^i(y) \psi(x)$, with $\psi(x)$ a 4D chiral
spinor. Setting $\varphi = U \phi$ we get the following equations of
motion in the SS basis
 \begin{equation}
 m \,  \phi^i -   \epsilon^{ij}  \, \frac{d \bar{\phi}_j}{dy} = 0 \, , \quad
 m \,  \bar{\phi}_j \, \epsilon^{ij} + \frac{d \phi^i}{dy} = 0 \; .
\label{bulkeom}
 \end{equation}
The parameter $m$ in Eq.~(\ref{bulkeom}) is the Majorana mass
eigenvalue of the 4D chiral spinor~\footnote{The bar acting on a
scalar quantity, as~e.g.~$\bar\phi_i$, and a chiral spinor,
as~e.g.~$\bar\psi$, denotes complex conjugation.}
\begin{equation}
i \sigma^\mu \de_\mu \bar{\psi} = m \, \psi \, , \qquad i
\bar{\sigma}^\mu \de_\mu \psi = m \, \bar{\psi} \quad .
\end{equation}
As a consequence of the transformation (\ref{cb}) the SS parameter
$\omega$ manifests itself only in the BC at $y=\pi R$~\footnote{Notice
that $U(y=0)=1$.  The roles of the branes and hence of $z_\pi $ and
$z_0$ can be interchanged by considering the SS transformation
$U'(y)\equiv U(y-\pi R)$.}:
\be \zeta_0\equiv\left.\frac{\phi^2}{\phi^1}\right|_{y=0}=z_0,\quad
\zeta_\pi\equiv\left.\frac{\phi^2}{\phi^1}\right|_{y=\pi R}=\frac{
\tan(\pi\omega) (i q_1 - q_2 - i q_3\, z_\pi) + z_\pi } {\tan(\pi\omega) (i q_1
\, z_\pi + q_2\, z_\pi + i q_3) + 1 }\;,
\label{SSbc}
\ee
where $\zeta_f$ are the BC's in the SS basis. In particular the BC
$\zeta_\pi$ is a function of $\omega$, $\vec q$ and $z_\pi$.  From
this it follows that we can always gauge away the SS parameter
$\omega$ in the bulk Lagrangian going into the SS basis through
(\ref{cb}). However now in the new basis $\omega$ reappears in one of
the BC's.

The bulk equations have the following generic solution
\begin{equation}
 \phi(y)=
  \begin{pmatrix}
    \bar a \cos(my)+\bar z_0 a \sin(my)
    \\
    -a\sin(my)+z_0 \bar a \cos(my)
  \end{pmatrix} \quad ,
 \label{sol}
\end{equation}
where $a$ is a complex number given in terms of $z_0$ and $\zeta_\pi$:
\begin{equation}
  a=\frac{z_0-\zeta_\pi}{|z_0-\zeta_\pi|}+\frac{1+z_0 \bar
  \zeta_\pi}{|1+z_0\bar \zeta_\pi |}\; .
\end{equation}
The solution (\ref{sol}) satisfies the BC's Eq.~(\ref{SSbc}) for the
following mass eigenvalues
\begin{equation}
m_n  = \frac{n}{R} + \frac{1}{ \pi
R }\,\arctan\left|\frac{z_0-\zeta_\pi}{1+z_0\,\bar \zeta_\pi}\right|\; ,
\label{masses}
\end{equation}
where $n\in \mathbb Z$. When $z_0=\zeta_\pi$ there is a zero mode and
this corresponds to an unbroken supersymmetry. Indeed, the only
sources of supersymmetry breaking reside on the branes (gaugino mass
terms) and setting them to cancel each other, $z_0=z_{\pi}$, preserves
supersymmetry~\cite{horava}. Once supersymmetry is further broken in
the bulk, an obvious way to restore it is by determining $z_\pi$ as a
function of $z_0$ and $\omega$ using the relation (\ref{SSbc}) with
$\zeta_\pi=z_0$.  This will lead to an $\omega$-dependent
brane-Lagrangian at $y=\pi R$. In this case we could say that
supersymmetry that was broken by BC's (SS twist) is {\it restored} by
the given SS twist (BC's)~\cite{SSBC}.

There is however a more interesting case: suppose the brane Lagrangian
determines $z_\pi$ to be
\begin{equation}
z_\pi=z(\vec q\,)\equiv\frac{\lambda-  q_3}{q_1 - i q_2}
\label{vH}.
\end{equation}
with $\lambda=\pm 1$.  This special value of $z_\pi$ is a fixed point
of the SS transformation, i.e.~$\zeta_f=z_f$.  For $z_\pi=z(\vec q\,)$
the spectrum becomes independent on $\omega$.  In other words, for
this special subset of boundary Lagrangians, the VEV of the field
$\vec q\cdot\vec V_5$ does not influence the spectrum. The reason for
this can be understood by going back to the Lagrangian which we used
to derive the BC's. From the relation (\ref{zBC}) one can see that
condition (\ref{vH}) is satisfied by the mass matrix
\begin{eqnarray}
M_{12}^{_{(\pi)}} &=& \lambda q_3 \nonumber \\
M_{11}^{_{(\pi)}} &=& -\lambda (q_1 + i q_2 ) \nonumber \\
M_{22}^{_{(\pi)}} &=& \lambda(q_1 -  i q_2 )
\label{msol}
\end{eqnarray}
which can be translated into a mass term at the boundary $y=y_\pi$
along the direction of the SS term, i.e.~$V^{(\pi)}=0$ and $T^{(\pi)}=
- \lambda\,\vec q\cdot\vec\sigma$ in the notation of
Eq.~(\ref{bmasses}). In particular this brane mass term preserves a
residual $U(1)_R$ aligned along the SS direction $\vec{q}$.  In other
words the SS-transformation $U$ leaves both brane Lagrangians
invariant and $\omega$ can be gauged away.  When we further impose
$z_0=z(\pm\vec q\,)$, i.e.~$V^{(0)}=0$ and $T^{(0)}=\pm T^{(\pi)}$ the
$U(1)_R$ symmetry is preserved by the bulk. In particular if
$z_0=z(\vec q\,)$ supersymmetry remains unbroken, although the VEV of
$\vec q\cdot \vec V_5$ is nonzero. One could say that in this case the
theory is persistently supersymmetric even in the presence of
the SS twist, with mass spectrum $m_n=n/R$. On the other hand if
$z_0=z(-\vec q\,)$ the theory is persistently
non-supersymmetric and independent on the SS twist: the mass spectrum
is given by $m_n=(n+1/2)/R$. In this case supersymmetry breaking
amounts to an extra $\mathbb Z_2^\prime$
orbifolding~\cite{Barbieri:2000vh}.

Actually (\ref{msol}) is the most general solution of
Eq.~(\ref{vH}). We can set $T^{(f)} = \vec{t_f} \cdot \vec \sigma$ and
$V^{(f)} = \vec{v_f} \cdot \vec \sigma$.  The constraint (\ref{det})
on the boundary mass-matrix translates into
\begin{equation}
{\vec{t_f}}^2 -  {\vec{v_f}}^2 = 1 \, ,   \qquad \vec{t_f} \cdot \vec{v_f} = 0
\quad .
\end{equation}
Consider now $f=\pi$, using an $SU(2)$ transformation we can always
rotate $\vec{v}_\pi$ in the $z$-direction. As a result without loss
of generality we can take $\vec{v}_\pi=(0,0,v_3)$.  Imposing that
$\zeta_\pi = z_\pi$ and ${\vec{t_f}}^2 - {\vec{v_f}}^2 = 1$ we get
\begin{equation}
\vec{t_f} = \lambda \, \vec q + \vec \theta(v_3) \, , \qquad \lambda^2
= \pm 1 \; ;
\end{equation}
where $\vec\theta$ is a vector which depends on $v_3$. The last
constraint $\vec{t}_f \cdot \vec{v}_f = 0$ is satisfied only if
$v_3=0$, which gives $\vec \theta(v_3=0) \equiv 0$. Then $ V^{(\pi)} =
0$ and $ T^{(\pi)} = - \lambda \, \vec q \cdot \vec \sigma $.

\section{\sc Super Yang-Mills action in the interval}
\label{SYM}

Up to now we have focused on the fermion sector spectrum. Adding the
complete vector multiplet does not invalidate our conditions for
supersymmetry restoration as long as the supersymmetry breaking brane
mass terms are of the form given by Eq.~(\ref{bmasses}). We would like
to show the invariance of the gaugino Lagrangian, Eq.~(\ref{total}),
under (global) supersymmetry.  To this end we will consider the Super
Yang-Mills multiplet containing the gauge field $B_M$ with field
strength $G_{MN}$, the gaugino $\Psi$, the real scalar $\Sigma$ and
the auxiliary $SU(2)_R$ triplet $\vec X$. Clearly since we are not
imposing a priori any BC on the fields in the action, we have to worry
about the total derivatives which arise in the variation of the bulk
action. The latter is given by
\begin{multline}
 S_{\rm bulk}^{\rm SYM}= \int_{\mathcal M}\biggl(
-\frac{1}{4}G_{MN}G^{MN} -\Sigma\, \mathcal D^2\,
\Sigma-\frac{1}{2}\mathcal D_M\Sigma \,\mathcal D^M\Sigma +2\,\vec
X\cdot\vec X\biggr.\\ \biggl.  +i\bar\Psi\gamma^M \mathcal D_M\Psi +ig
f_{ABC}\bar\Psi^A\Psi^B\Sigma^C\biggr).
\label{globalbulk}
\end{multline}
Here the sum over the adjoint indices of the fields is suppressed and
$\mathcal D$ denotes the gauge covariant derivative.  Under a global
supersymmetric transformation the Lagrangian transforms into a total
derivative giving rise to the supersymmetry boundary-variation:
\be
\delta_\epsilon S_{\rm bulk}^{{\rm SYM}}=\int_{\cal\partial M}\bar
\epsilon i\gamma^5\rho ,
\ee
where $\rho$ is given by
\be
\rho=\left(i\vec X \cdot \vec\sigma -\Sigma
\, \gamma^M \mathcal D_M-\frac{1}{4}\,\gamma^{MN}G_{MN}
-\frac{1}{2}\, \gamma^M \mathcal D_M\Sigma\right)\Psi .
\ee
To compensate for this we add the brane action
\be
S_{\rm brane}^{{\rm SYM}}=\int_{\cal\partial M}\left( 2 \vec T^{(f)}
\cdot\Sigma\vec X+\frac{1}{2}\bar\Psi T^{(f)}\Psi\right)
\label{globalbrane}
\ee
which transforms into
\be
\delta_\epsilon S_{\rm brane}^{{\rm SYM}}=\int_{\cal\partial
  M}\bar \epsilon\, T^{(f)}\rho .
\ee
Now the supersymmetry variation at each boundary is proportional to
$(1+i\gamma^5 T^{(f)})\epsilon(y_f)$.  Denoting with $\xi$ [see
Eq.~(\ref{uno})] the upper part of $\epsilon$, whenever $(\vec
T^{(f)})^2=1$ these variations can cancel provided the transformation
parameter satisfies the BC's $\xi^2=z(\vec T^{(f)})\,\xi^1$.  The only
possibility is that $T^{(0)}=T^{(\pi)}$, since $\epsilon$ is constant
for global supersymmetry.  Notice that according to Eqs.~(\ref{det})
and (\ref{TVm}), this gives rise to the same BC's for the gaugino,
$\eta^2=z(\vec T^{(f)})\,\eta^1$.  The remaining EOM then fix the BC's
$G_{\mu 5}=\vec X=\Sigma=0$. The bottom line of the off-shell approach
is that, in the presence of a boundary, at most one supersymmetry can
be preserved.  Global SUSY invariance for the action of a vector
multiplet singles out a special boundary mass term for gauginos such
that $z_0 = z_\pi$ which is at origin of the zero mode in the spectrum
[see Eq.~(\ref{masses}) for $\omega =0$~\footnote{In the global theory
on the interval all supersymmetry breaking is encoded in the $T^{(f)}$
mass matrix. There is no auxiliary field $V_M$ whose VEV could
contribute to the breaking, nor can one choose a SS twist by hand,
since the BC's are uniquely fixed by the equations of motion.}].  We
expect there to be a locally supersymmetric extension of the action
(\ref{globalbulk})+(\ref{globalbrane}) for $T^{(0)}\neq T^{(\pi)}$.
In this case the $SU(2)_R$ auxiliary gauge connection $\vec V_M$ from
the supergravity multiplet gives an additional source of supersymmetry
breaking.  Notice that for a globally supersymmetric vacuum there must
then be a solution to the Killing spinor equation
\be
\gamma^5D_5 \epsilon(y)=0,\qquad \xi^2(y_f)=z(\vec
T^{(f)})\,\xi^1(y_f),
\ee
where $D_M$ is covariant with respect to $SU(2)_R$. These equations
coincide with the zero mode condition for the gaugino considered
above.

\section{\sc Fermions in the orbifold}
\label{orbifold}

In this section we will consider the same system of a symplectic
Majorana spinor but in the more common orbifold approach. As we have
seen in the previous section, in the interval framework once the
action is given we obtain the bulk equations supplemented with a
consistent set of BC's.  Let us now study what happens in the orbifold
geometry for an $SU(2)_R$ fermionic doublet $\Psi^i$ where $SU(2)_R$
is identified with the automorphism of $N=2$ supersymmetry algebra.
We will consider the direct product of a flat 4D Minkowski spacetime
times the orbifold $S^1/\mathbb{Z}_2$, the (flat) geometry considered
in section~\ref{interval}.  In the $S^1/\mathbb{Z}_2$ orbifold the
fifth coordinate runs now along the circle $y\in[-\pi R,\pi R]$ and we
assign to the spinors the following parities
\begin{equation}
\begin{split}
  &\eta^1(x, \, -y) = \eta^1(x,y) \, , \qquad
  \eta^2(x, \, -y) = - \eta^2(x,y),\\
  &\eta^1(x, \, \pi R-y) = \sigma\,\eta^1(x,\pi R+y) \, , \qquad \eta^2(x,
  \, \pi R-y) = - \sigma\,\eta^2(x,\pi R+y),
\end{split}
\label{parities}
\end{equation}
where $\sigma=\pm 1$. The second condition is often replaced by
demanding periodicity ($\sigma=+1$) or anti-periodicity ($\sigma=-1$)
and corresponds to an intrinsic parity for the inversion with respect
to $y=\pi R$.  The orbifold Lagrangian is then given by
\begin{equation}
{\cal L} = i \bar{\Psi} \gamma^{{}_M} D_{_M} \Psi + 2\,\delta(y)
\left( N_{ij}^{_{(0)}} \, \eta^i \, \eta^j + \, \text{h.c.} \right) +
2\,\delta(y- \pi R) \left( N_{ij}^{_{(\pi)}} \, \eta^i \, \eta^j + \,
\text{h.c.} \right) \; .
\label{lagorb}
\end{equation}

Dirac-like mass terms mixing $\eta_1$ and $\eta_2$,
$N_{12}^{_{(0,\pi)}}$, must have an odd profile as it is obvious from
the parity assignments in Eq.~(\ref{parities}). If they are continuous
at $y=0,\pi R$ they do not contribute to the brane mass Lagrangian in
(\ref{lagorb}); therefore if they contribute they must possess a
discontinuity at the fixed points $y=0,\pi R$. The simplest ansatz is
that near the fixed point at $y=y_f$ ($f=0,\pi$) they behave
as~\footnote{Of course different ans\"atze could be considered, as an
odd power of $\epsilon(y)$ or any other odd function. We will just
consider in this paper the usual case of a linear behavior in
$\epsilon(y)$.}
\begin{equation}
N_{12}^{_{(f)}}=N_D^{_{(f)}}\epsilon_f(y)
\label{Diracmas}
\end{equation}
where $\epsilon_0=\epsilon(y)$, $\epsilon_\pi=\epsilon(\pi R-y)$ and
the sign function $\epsilon(y)$ is defined as
\begin{equation}
\epsilon(y)=2\int_0^y \delta(z)dz\ .
\label{epsilon}
\end{equation}
In the simplest case where there is no mass localized at the orbifold
fixed points (e.g. $N^{_{(f)}} = 0$) there is a straightforward
correspondence between the orbifold and the interval approach: we can
take the fields continuous across the orbifold fixed points. Then
using the parity assignment (\ref{parities}) and periodicity we have
\begin{equation}
\eta^2(0^+) = 0 \, , \qquad \eta^2({\pi R}^-) = 0 \quad .
\label{sobc}
\end{equation}
Being the EOM the same we recover the interval result simply using as
BC $z_f =0$. As a result, when no mass term is present at the orbifold
fixed points the parity assignment in the orbifold is equivalent to
the choice of BC's in the interval.  As soon as we turn on a mass term
in the orbifold fixed point the correspondence is much more
involved. First of all we have to immediately face a technical problem
inherent to the orbifold construction: to give a meaning to the fixed
point Lagrangian which contains the product of distributions with
overlapping singularities. In order to do that often implicit or
explicit assumptions regarding the continuity properties of the fields
are made. We want to stress here that such assumptions can lead to
inconsistencies since solutions to the EOM might not exist. On the
other hand by regularizing the delta functions in (\ref{lagorb}) one
can consistently assume all fields to be smooth while only in the
limit of a sharp delta the solutions to the EOM will develop
discontinuities. For a ``sharp'' delta function, $\epsilon(y)$ is
simply the step function with value $+1$ ($-1$) for $y>0$ ($y<0$).
For a ``regularized'' delta function $\epsilon(y)$ is a regular odd
function that therefore satisfies the property $\epsilon(0)=0$. In all
the expressions that follow delta and epsilon functions should be
considered as regularized with the sharp limit implicitly taken at the
end of the calculation.

Our strategy will thus be the following. In a first step we solve the
EOM close to the branes using the orbifold BC's
(\ref{parities}). Taking the limit of sharp distributions we determine
the precise jump profile across the branes which will result in
modified BC's, at $0^+$ and $\pi^-$ respectively. These modified BC's
can again be encoded in complex numbers denoted by $z_{0^+}$,
$z_{\pi^-}$ which are functions of the brane masses. Since the bulk
EOM are identical to those in the interval approach we can directly
use the results from section~\ref{interval}, that is Eqs.~(\ref{sol})
to (\ref{masses}), with the only replacements $z_0\to z_{0^+}$ and
$z_\pi\to z_{\pi^-}$. This establishes the precise relation between
the orbifold and interval approaches.

We use the notation of section~\ref{interval}, where
$\eta^i(x,y)=\varphi^i(y)\psi(x)$ are the components of the
symplectic-Majorana spinors in the Hosotani basis. Going to the SS
basis, $\Phi^i$, $ \Psi = \exp{\left(-i\, \vec{q}\,\vec{\sigma}\,
\omega\, y/R\right)} \, \Phi$ we obtain the following EOM in the
orbifold geometry
\begin{gather}
\varepsilon^{ij} \frac{d \phi_j }{dy} + m \, \bar{\phi}^i -2
\sum_{f=0,\pi} \delta(y-y_f) \, N_{ij}^{(f)} \, \phi^j =0
\label{oe}
\end{gather}
To determine the precise form of the discontinuities we consider the
EOM close to the fixed points at $y=0$
\begin{gather}
\frac{d \varphi^1}{dy} + 2 \left( N^{_{(0)}}_{22}
\varphi^2+N^{_{(0)}}_{12} \varphi^1\right)\,
\delta(y)=0\label{oe1b}\\
-\frac{d \varphi^2}{dy} +2 \left(
N^{_{(0)}}_{11}\varphi^1+N^{_{(0)}}_{12}\varphi^2\right) \,
\delta(y)= 0\;
\label{oe2b}
\end{gather}
Note that we have neglected the term $m \bar \phi^i$ as well as the
one $\propto \delta(y-\pi)$ since they are negligible close to
$y=0$. This approximation becomes more and more accurate the sharper
the distributions are taken, and it is in fact exact in the singular
limit.  At $y=\pi R$ the equations are in the same form {\it mutatis
mutandi} $0$ by $\pi$.  From here it is easy to see that making
assumptions about continuity of the fields may fail. For instance,
continuity at $y=0$~\footnote{To see this one typically integrates
between $0$ and $0^+$. Note that a discontinuous even function has the
same value at $0^-$ and $0^+$ but a different one at $0$.} of both
$\varphi_1$ and $\varphi_2$ is clearly consistent only with
$N^{_{(0)}}_{ij}=0$, while the weaker assumption of only $\varphi_1$
smooth is inconsistent unless $N^{_{(0)}}_{22}=N^{_{(0)}}_{12}=0$ or
$N^{_{(0)}}_{11}=N^{_{(0)}}_{12}=0$.  To avoid these difficulties we
proceed as follows.  We can solve Eqs.~(\ref{oe1b}) and (\ref{oe2b})
by assuming (close to the fixed point at $y=0$) the functional
dependence $\varphi^i=\varphi^i(\epsilon(y))$ and using the chain rule
as
\begin{equation}
\frac{d\varphi^i}{dy}=2\,\left(\varphi^i\right)^\prime\,\delta(y)
\label{chain}
\end{equation}
where $\left(\varphi^i\right)^\prime\equiv d\varphi^i/d\epsilon$.
Using now (\ref{chain}) we can cast Eqs.~(\ref{oe1b}) and (\ref{oe2b})
as
\begin{gather}
 \left(\varphi^1\right)^\prime +  N^{_{(0)}}_{22}
\varphi^2+  N^{_{(0)}}_{12} \varphi^1=0\label{oe1d}\\
 \left(\varphi^2\right)^\prime-
N^{_{(0)}}_{11}\varphi^1- N^{_{(0)}}_{12}\varphi^2= 0
\label{oe2d}
\end{gather}
Passing from Eqs.~(\ref{oe1b}) and (\ref{oe2b}) to Eqs.~(\ref{oe1d})
and (\ref{oe2d}) makes sense for any regularized delta function for
which we can solve our EOM. As we said we can consistently take the
sharp limit after solving the EOM. In that case, using $\epsilon(0)=0$
we can solve Eqs.~(\ref{oe1d}) and (\ref{oe2d}) with the BC's
$\varphi^2(0)=0$, $\varphi^1(0)=1$. One gets for any regularized delta
function the {\it exact} analytical solution
\begin{gather}
\varphi^1(y)= e^{{\displaystyle \frac{-N^{_{(0)}}_D}{2}\epsilon^2(y)} }
 \ {_1}F_1\left[\frac{-N^{_{(0)}}_{11}N^{_{(0)}}_{22}}{4\,N^{_{(0)}}_D};
\frac{1}{2};N^{_{(0)}}_D\epsilon^2(y)\right]\label{sol1}\\
\varphi^2(y)=N^{_{(0)}}_{11}\,\epsilon(y)\,
 e^{{\displaystyle \frac{-N^{_{(0)}}_D}{2} \epsilon^2(y)}}
 \ {_1}F_1\left[1-\frac{N^{_{(0)}}_{11}N^{_{(0)}}_{22}}{4\,N^{_{(0)}}_D};
\frac{3}{2};N^{_{(0)}}_D\epsilon^2(y)\right]\label{sol2}
\end{gather}
where $_1F_1$ is the Kummer confluent hypergeometric function~\footnote{The
Kummer confluent hypergeometric function is defined by the series
expansion
$${\displaystyle
_1F_1[a;b;z]=1+\sum_{k=1}^\infty\frac{a(a+1)\cdots
(a+k-1)}{b(b+1)\cdots (b+k-1)}\, \frac{z^k}{k!} }\ .
\label{kummer}
$$
}. 
From Eq.~(\ref{sol1}) and the definition of the Kummer function
it is easy to check that the odd field is discontinuous
at the brane unless $N^{(0)}_{11}=0$.  The even field makes a jump
unless $N^{(0)}_{D}=0$ and $N^{(0)}_{11}\,N^{(0)}_{22}=0$. In fact the
solutions (\ref{sol1})-(\ref{sol2}) have an interesting expression in
the limit where the Dirac mass $N^{(0)}_{D}\to 0$. They are given by
\begin{gather}
\lim_{N^{(0)}_{D}\to 0}\varphi^1(y)=\cos\left(\sqrt{
N^{(0)}_{11}\,N^{(0)}_{22}}\,\epsilon(y)\right)\label{sol10}\\
\lim_{N^{(0)}_{D}\to
0}\varphi^2(y)=-\sqrt{\frac{N^{(0)}_{11}}{N^{(0)}_{22}}}\,\sin\left(\sqrt{
N^{(0)}_{11}\,N^{(0)}_{22}}\,\epsilon(y)\right)\label{sol20}
\end{gather}

Similarly we can solve the Eqs.~(\ref{oe}) on the vicinity of $\pi R$
by assuming that near the brane there is a functional dependence as
$\varphi^i=\varphi^i(\epsilon_\pi(y))$ and using
$d\epsilon_\pi(y)/dy=-2\delta(y-\pi R)$. The solutions can be easily
read off from Eqs.~(\ref{sol10}) and (\ref{sol20}) by simply changing
$ N^{(0)}_{ij} \rightarrow - N^{(\pi)}_{ij}$ and
$\varepsilon_0\to\varepsilon_\pi$ (for $\sigma = 1$) or by changing $
N^{(0)}_{ij} \rightarrow N^{(\pi)}_{ij}$,
$\varepsilon_0\to\varepsilon_\pi$ and
$\varphi^1(y)\leftrightarrow\varphi^2(y)$, i.e.~by simply label
changing $1\leftrightarrow 2$ and $0\to\pi$ (for $\sigma=-1$).

We can now use the behaviour of the solutions close to the fixed
points to read out the jumps for odd and even fields caused by the
presence of the delta-functions in the EOM.
We have at $y=0^+$
\be
\varphi^2(0^+)=N^\zero_{11}\,
\frac{{_1}F_1\left[1-
{\displaystyle\frac{N^{_{(0)}}_{11}N^{_{(0)}}_{22}}{4\,N^{_{(0)}}_D} };
\frac{3}{2};N^{_{(0)}}_D\right]}{{_1}F_1\left[{\displaystyle
\frac{-N^{_{(0)}}_{11}N^{_{(0)}}_{22}}{4\,N^{_{(0)}}_D} };
\frac{1}{2};N^{_{(0)}}_D\right]}\,\varphi^1(0^+) \ .
\ee
Similar expressions at $\pi^-$ are obtained by doing the
before-mentioned changes.  Now we can identify the values of the $z_f$
parameters entering in the mass-formula. From the jumps at $y=0^+$ one
obtains:
\begin{gather}
z_{0^+}=N^\zero_{11}\,
\frac{{_1}F_1\left[1-{\displaystyle
\frac{N^{_{(0)}}_{11}N^{_{(0)}}_{22}}{4\,N^{_{(0)}}_D} };
\frac{3}{2};N^{_{(0)}}_D\right]}{{_1}F_1\left[{\displaystyle
\frac{-N^{_{(0)}}_{11}N^{_{(0)}}_{22}}{4\,N^{_{(0)}}_D} };
\frac{1}{2};N^{_{(0)}}_D\right]} \; ,
\label{zorb}
\end{gather}
while $z_{\pi^-}$ can be obtained by the above mentioned trivial
substitutions.  In the particularly simple case of no Dirac mass one
obtains the expressions
\begin{gather}
\lim_{N^\zero_D\to 0} z_{0^+}=-\sqrt{\frac{N^\zero_{11}}{N^\zero_{22}}}
\tan\left(\sqrt{N^\zero_{11}N^\zero_{22}}\right),
\label{zorb0} \quad ;\\
\lim_{N^{_{(\pi)}}_D\to 0} z_{\pi^-}=i\sqrt{\frac{N^\ppi_{11}}{N^\ppi_{22}}}
\left[i\tan\left(\sqrt{N^\ppi_{11}N^\ppi_{22}}\right)\right]^\sigma\ .
\label{vorb0}
\end{gather}

As for the solution of the EOM far from the branes, the form is
identical to the one in the interval case, Eq.~(\ref{sol}), but now in
the mass formula one has to use the quantities $z_{0^+}$ and
$z_{\pi^-}$, which are given by different functions of the brane
masses.

To cross-check the validity of this procedure we have done a
numerical integration of the full EOM, Eqs.~(\ref{oe}). We indeed find
that solutions only exist for the mass eigenvalues given by
Eq.~(\ref{masses}), where $z_f$ are calculated from the mass
parameters through Eqs.~(\ref{zorb}).

Let us summarize the orbifold approach. In this framework the BC's
are uniquely fixed by the parity assignments, while the boundary
Lagrangian cause the wavefunctions to jump. We have shown how these
jumps can be computed by regularizing the delta functions, thereby
obtaining smooth wave-functions. The results are manifestly
independent of the regularization, so this gives a well defined
procedure. The jumps define new BC's at $0^+$ and $\pi^-$ which can be
used to solve the bulk equations in the interval $[0^+,\pi^-]$. If
these BC's are aligned with the Scherk-Schwarz direction, we again
find the spectrum to be independent of the SS parameter.

Finally let us stress that we have computed the orbifold mass
eigenvalues and eigenfunctions for arbitrary brane mass terms. Our
results generalize (and agree with) the particular cases previously
considered in the literature~\cite{Bagger:2001qi,Delgado:2002xf,SSBC}.

\section{\sc Warped geometry}
\label{warped}

In this section we extend our previous results in the interval
approach to warped geometry.  The $SU(2)_R$ automorphism group found
in the off-shell bulk supergravity action in flat space is broken by
the warping down to two surviving $U(1)$'s. One of them is gauged by
the graviphoton $A_M$ with gauge coupling $g$, and the other $U(1)$
invariance becomes redundant after the elimination of the auxiliary
field $\vec{V}$ in terms of the graviphoton.  The gauged $U(1)$
couples to the fermions via the off shell Lagrangian~\cite{zucker}
\be
{\cal L}_{\text{bulk}} = \frac{i}{2} \bar{\Psi} \gamma^{{}_M}
D_{{}_M} \Psi - \frac{i}{2} D_{{}_M} \bar{\Psi} \gamma^{{}_M}\Psi
-2 c \bar\Psi\,\vec t \cdot\vec\sigma\, \Psi \; ;
\ee
where $\Psi$ is a symplectic-Majorana fermion, the field $\vec t$ is
an $SU(2)_R$-triplet auxiliary-field from the minimal supergravity
multiplet, the covariant derivative is given by
\begin{equation}
D_{{}_M} = \de_{{}_M} - \frac{1}{4} \omega_{{}_{AB \, M}} \,
\gamma^{{}_{AB}} - \frac{i}{2}\, \vec{\sigma}\cdot \vec V_M\quad .
\end{equation}
and $c$ is the bulk mass term fixed by supersymmtery in the warped
space. In particular $c=1$ for gauginos.

The equations of motion then imply that~\cite{zucker}
\be
\vec t=\frac{1}{4\sqrt 3} g \,\vec q \; ;
\ee
while for $V_M$ one finds a relation with the graviphoton $A_M$ as
\be
d(\vec q\cdot\vec V-2gA)=0,\qquad \vec q^{\,\,2}=1 \; .
\ee
which gives the solution
\be
\vec V_M =2\, \vec q \,(C_M+g\, A_M) \qquad dC=0 \; .
\label{vevV}
\ee
The one form $C$ is closed, $dC=0$, but not necessarily trivial,
$C=0$.  For nonzero $g$ it can however be absorbed into the
graviphoton $A_M$ by a shift~\footnote{Note that this does not happen
in the flat case.}. All the Scherk-Schwarz breaking will then be
encoded in the graviphoton VEV, $\langle A\rangle=\omega\, dy$.  The
graviphoton will gauge the $U(1)_R\subset SU(2)_R$ subgroup defined by
the unit vector $\vec q$.

For the on-shell gaugino action we thus get
\be
{\cal L}_{\text{bulk}} = \frac{i}{2} \bar{\Psi} \gamma^{{}_M}
D_{{}_M} \Psi - \frac{i}{2} D_{{}_M} \bar{\Psi} \gamma^{{}_M}\Psi
-c \, \frac{g}{2\sqrt 3}\bar\Psi\,\vec q\cdot\vec\sigma\, \Psi \; ;
\ee
where the covariant derivative is now given by
\begin{equation}
D_{{}_M} = \de_{{}_M} - \frac{1}{4} \omega_{{}_{AB \, M}} \,
\gamma^{{}_{AB}} - i\, \vec{q} \cdot \vec{\sigma} \,
g A_M \ .
\end{equation}

Let us now turn to the bosonic sector.  In order to obtain a viable
phenomenology we will restrict our discussion to the Randall-Sundrum
model~\cite{RS}, which leads to $4D$ Minkowski spacetime. The warping
in the bulk must then be balanced with the brane tensions ending up
with the tunings of the Randall-Sundrum model. Moreover local
supersymmetry requires the bulk (AdS) cosmological constant to be
related with the fermion coupling to the graviphoton,
$\Lambda=-g^2$. The metric in conformal coordinates can be written as
\begin{equation}
ds^2 = a^2 \left[ \eta_{\mu \nu} dx^\mu dx^\nu + du^2 \right]\, ,
\qquad a(u) = (k u)^{-1} \; .
\end{equation}
The conformal coordinates $(x,u)$ are related with the coordinates
in~\cite{RS} by $u(y) = k^{-1} \exp(k y)$.  The constant $k$ is
related to the gauge coupling $g$ as $g=\sqrt 3 k$.  The graviphoton
background in the conformal frame reads as
\be
A=\omega\, a\, du  \; .
\ee
Concerning the boundary Lagrangian $T^{{}_{(0)}}, \, V^{{}_{(0)}}$
will be the mass terms in the UV brane located at $u =
u_{{}_{UV}}=1/k$ and $(u_{{}_{IR}} k)^5 \, T^{{}_{(\pi)}}, \,
(u_{{}_{IR}} k)^5 \, V^{{}_{(\pi)}}$ the mass terms in the IR brane

Proceeding as in section~\ref{interval}, the BC at the IR brane is
twisted by $\omega$ as given in (\ref{SSbc}).  Using the same notation
and setting $\hat \omega = \omega/k$, the bulk Dirac equation can be
written in the SS frame as
\begin{equation}
i \gamma^{{}_M} \de_{{}_M} \left(a^{2} \, \Phi \right ) + \frac{1}{2}
\, c \, k a^{3} \,
\vec{q} \cdot \vec{\sigma} \Phi = 0
\quad .
\label{wdirac}
\end{equation}
The massive KK modes associated with the solutions of (\ref{wdirac})
can be expressed as combinations of Bessel functions of order $\nu =
|c \pm 1|/2$, for $\vec q=(0,0,1)$, where the two signs refer to the
two gaugino components. Let us focus on the zero mode, its
wavefunction is given by
\begin{equation}
\varphi(u) = a^{-2}(u) \exp\{- \vec{q} \cdot \vec{\sigma} \log(k
u)(i\hat{\omega}+ c/2 )\} \, \varphi(k^{-1}) \quad .
\end{equation}
The condition for the existence of a zero mode coming from enforcing
the BC's is the following
\begin{equation}
\begin{split}
&(z_0 - \zeta_\pi) \, \cosh \theta + \left[-q_1 (z_0 \, \zeta_\pi -1)
+ i \, q_2 \, (1+z_0 \, \zeta_\pi) - q_3 \, (z_0 + \zeta_\pi) \right]
\, \sinh \theta=0 \; ; \\ &\theta = \frac{1}{2}\log
\left(u_{{}_{IR}}/u_{{}_{UV}} \right)=\frac{k\pi}{2} \; .
\end{split}
\label{warpzero}
\end{equation}

In the absence of any fine tuning, the coefficients of $\sinh \theta$
and $\cosh \theta$ should vanish separately and we get
\begin{equation}
z_0 = \zeta_\pi = \frac{\lambda-  q_3}{q_1 - i q_2} \equiv z(\vec{q}) \, 
\label{vH2}
\; ;
\end{equation}
the same condition as in flat space. There is however an important
difference: the values of $z_f$ are fixed by imposing that the SUSY
transformations are fulfilled 
\be \left(z_f \, \delta_\epsilon \, \eta^1 \, -
\, \delta_\epsilon \, \eta^2 \right)|_{y_f} = 0 \; .
\label{sss}
\ee
This will lead $z_0, z_{\pi}$ to depend on the bosonic tensions on the
respective branes $\tau_0, \tau_{\pi}$ and on the bulk cosmological
constant $g$. When further imposing one of the Randall-Sundrum
tunings, $\tau_0=\tau_{\pi}$, one gets $z_0=z_{\pi}$, and when
enforcing the brane-bulk balance, $\tau_0 = \sqrt{24}g $, one obtains
$z_0=z_\pi=z(\vec q)$~\footnote{The quantity $z(\vec q)$ was defined
in Eq.~(\ref{vH}).}. Therefore, persistent supersymmetry is
guaranteed.  Any other BC's deviating from this value would explicitly
violate local supersymmetry~\footnote{The special case of flipped
BC's~\cite{Gherghetta:2000kr} corresponds to $z_0=z(\vec q)$,
$z_\pi=\zeta_\pi=z(-\vec q)$.}~\cite{Gherghetta:2000kr,bagger1}. This
is in contrast to the flat case, where any BC's for the fermions
respect local supersymmetry, and the breaking of $N=1$ SUSY by
non-aligned BC's is spontaneous.  On the other hand in the warped
case, imposing local supersymmetry implies the BC's~(\ref{vH2}) and
supersymmetry is persistent whatever the VEV of $\vec q \cdot \vec
V_M$, Eq.~(\ref{vevV}), is. This fact was already noticed in
Refs.~\cite{Pomarol,bagger2,lalak,Hall:2003yc}.  In our formalism this
persistence is a consequence of the alignment of bulk and brane
breaking. This alignment is due to the relation between bosonic
stability in the metric and the fermionic sector via supersymmetry
transformations, Eq.~(\ref{sss}).

\section{\sc Conclusions}
\label{conclusions}

We have presented in this paper a detailed study of supersymmetry
breaking and restoration when two sources of breaking are present,
both Scherk-Schwarz and boundary mass terms. The mass spectrum of
fermions is obtained through the equations of motion in the bulk and
brane boundary conditions. These boundary conditions are extracted
either in the interval approach, by extremizing the boundary action,
or in the orbifold language, by assigning parities to the fields.  We
compare the two approaches and compute the nontrivial jump profiles of
the wavefunctions in the orbifold picture for general brane mass
terms. With our procedure of dealing with fermions we study in both
approaches supersymmetry breaking by adding the Scherk-Schwarz
breaking terms and computing the mass spectrum. We find out that for a
suitable tuning of the boundary actions supersymmetry is always
present for arbitrary values of the Scherk-Schwarz parameter.  As an
application of the interval formalism, we construct bulk and boundary
actions for supersymmetric Yang-Mills theories. Finally we extend our
results to the warped Randall-Sundrum background.  In this case 
the SS-direction coincides with the gauged $U(1)_R$ subgroup, and
local supersymmetry enforces alignment of the boundary conditions with
the latter. Therefore any misalignment is an explicit form of
supersymmetry breaking and spontaneous SS breaking becomes
impossible. This conclusion agrees with the findings in
Refs.~\cite{Pomarol,bagger2,lalak,Hall:2003yc}.  Within our formalism,
it follows from the alignment of bulk and brane breakings once
one relates the Randall-Sundrum tunings in the bosonic sector with the
fermionic boundary conditions.


\vspace*{7mm}
\subsection*{\sc Acknowledgments}

\noindent This work was supported in part by the RTN European Programs
HPRN-CT-2000-00148 and HPRN-CT-2000-00152, and by CICYT, Spain, under
contracts FPA 2001-1806 and FPA 2002-00748 and grant number
INFN04-02. One of us (V.S.) thanks T.~Okui and J.~Hirn for useful
discussions and to the Particle Theory Department at Boston University
and the IPPP Physics Department at Durham University, where part of
this work has been done. Another of us (G.v.G.) thanks the Theory
Division of the Deutsches Elektronen-Synchroton (DESY) where part of
this work was done.  Finally three of us (L.P., A.R. and V.S.) would
like to thank the Theory Department of IFAE, where also part of this
work has been done, for hospitality.


\end{document}